\documentstyle[multicol,aps,prb,epsfig]{revtex}

\begin{document}
\pagestyle{empty}

\newcommand{\bc}{\begin{center}}
\newcommand{\ec}{\end{center}}
\newcommand{\be}{\begin{equation}}
\newcommand{\ee}{\end{equation}}
\newcommand{\beqn}{\begin{eqnarray}}
\newcommand{\eeqn}{\end{eqnarray}}
\newcommand{\siml}{\stackrel{<}{\sim}}
\newcommand{\simg}{\stackrel{>}{\sim}}

\begin{multicols}{2}
\narrowtext
\parskip=0cm

\noindent
{\large\bf Comment on "Liquid Limits: The Glass Transition and Liquid-Gas 
Spinodal Boundaries of Metastable Liquids"}
\smallskip

In a recent paper \cite{Sastry}, Sastry studies by numerical simulations
the phase diagram of a simple fragile glass-forming liquid, the binary 
mixture with Lennard-Jones (LJ) interactions introduced by Kob and Andersen, 
presenting very interesting and clear results. Among other 
things, he obtains the position of the liquid-glass transition 
temperature as function of density $\rho$. He studies both the 
temperature $T_0(\rho)$ where the diffusion coefficient (of $A$ particles)
drops to zero according to the Vogel-Fulcher-Tammann-Hesse law,
and the temperature where the extrapolation of data on the configurational 
entropy $S_c(\rho,T)$ vanishes, $S_c$ being evaluated as the difference 
between the liquid entropy and the harmonic amorphous solid entropy. 
These two ways give values remarkably close.

We apply to the LJ binary mixture, at various density values, the
analytic approach to structural glass thermodynamics recently
introduced \cite{MePa}, which allows to obtain both the liquid-glass
transition temperature $T_K$ and properties in the glassy phase below
$T_K$ doing a first principles computation. In one version of this
approach, which implements analitically the proposal of inherent
structure dominance at low temperature, one needs to compute the liquid
free-energy density $f$ and two-point correlation functions $g(r)$ at
$T \simg T_K$.

The theory and different ways for implementing it are discussed in detail 
elsewhere \cite{MePa,Me,CoMePaVe,CoPaVe}. We only recall 
that, within this framework,
at $T_K$ the system undergoes a second order transition corresponding to a 
one step replica simmetry breaking. As one could expect 
intuitively, it is found that in the region $T\simg T_K$, where the phase 
space can be partitioned in `valleys', the entropy of the liquid corresponds
to the entropy of a typical valley $S_{sol}$ plus the configurational 
entropy $S_c$, and the Kauzmann temperature is defined by the condition 
$S_c(T_K)=0$.

We use the harmonic approximation scheme, which gives a simple
formula for the single valley entropy in the liquid phase:
$$
S_{sol}=\frac{3 N}{2} \ln \left (2 \pi \: e \: T \right) - 
\frac{1}{2} \langle {\mbox Tr} \ln {\cal M} \rangle,
$$
where ${\cal M}$ is the $3 N \times 3 N$ Hessian (i.e. the matrix
of second derivative of the potential with respect to the $N \: d$ particle 
coordinates) and $\langle \cdot \rangle$ means the Boltzmann average.
We are therefore approximating $S_{sol}$ by only keeping the harmonic 
vibrational contribution. Further approximations are needed for computing 
the spectrum.
Here we consider fluctuations of the diagonal part of ${\cal M}$ up to second 
order \cite{CoPaVe}, while the whole nondiagonal contribution is summed up 
by means of a kind of chain approximation.

In the computation of the liquid free energy and of the correlation
functions in the high density region we use the Hypernetted Chain 
Approximation (HNC). Unfortunately this approximation overestimates the
gas-liquid transition temperature and a more refined one is necessary
at low densities. Liquid quantities at $\rho=1.2$ have been computed
\cite{CoPaVe} within a scheme similar to that introduced by Zerah and
Hansen (ZH), which also has the advantage of reducing the error in the
approximation (of order $10 \%$ in the HNC case).

\begin{figure}
\epsfig{figure=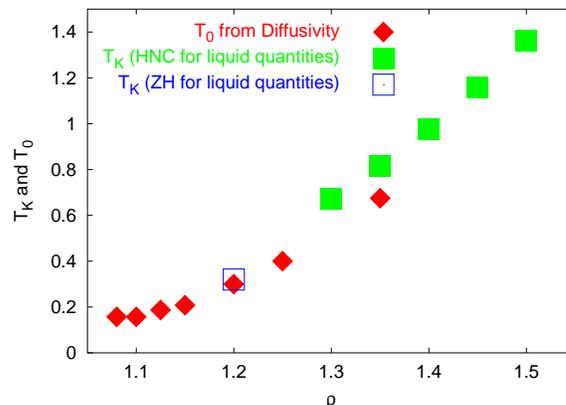,angle=270,width=8cm}
\caption{The data on $T_0$ obtained by Sastry \cite{Sastry} compared with
our analytical results.}

\end{figure}

We plot in [Fig. 1] the theoretical predictions on $T_K$ at different density 
values, compared with the data on $T_0$ by Sastry. Despite of the error in
liquid quantities obtained by HNC, there is a quantitative agreement which 
makes evident that the analytic approach works well and it can be used 
as a powerful tool for a better understanding of glassy systems.

To conclude, we stress that, at least in this case, the
Kauzmann temperature turns out to be directly comparable with the temperature
where the diffusion drops to zero, which strongly supports the theoretical 
expectation of an effective ergodicity breaking.

\bigskip
\noindent
B. Coluzzi$^{(a)}$, G. Parisi$^{(a)}$ and P. Verrocchio$^{(b)}$

{\small

a) Dipartimento di Fisica, INFN and INFM, Universit\`a di Roma
{\em La Sapienza}, P.le A. Moro 2, 00185 Roma, Italy.\\

\vskip-0.3cm
b) Dipartimento di Fisica, Universit\`a di Trento, I-3805
Povo, Trento, Italy.
}
\bigskip


\end{multicols}

\end{document}